\documentstyle[11pt,aasms4,rotating]{article}

\def\simlt{\lower.5ex\hbox{$\; \buildrel < \over \sim \;$}}
\def\simgt{\lower.5ex\hbox{$\; \buildrel > \over \sim \;$}}

\def\gcm3{{\rm\,g\,cm^{-3}}}
\def\ncm3{{\rm\,cm^{-3}}}

\def\>{$>$}
\def\<{$<$}

\def\refbook#1{\refindent#1}
\def\refindent{\par\noindent\hangindent=3pc\hangafter=1 }
\def\aa#1#2#3{\refindent#1, A\&A, {\bf#2}, #3.}

\def\apj#1#2#3{\refindent#1, {\it ApJ}, {\bf#2}, #3.}
\def\apjlett#1#2#3{\refindent#1, {\it ApJ (Letters)}, {\bf #2}, #3.}
\def\apjsup#1#2#3{\refindent#1, ApJS, #2, #3}

\def\nature#1#2#3{\refindent#1, {\it Nature}, {\bf #2}, #3.}

\lefthead{Melia et al.}
\righthead{The Broad-band Spectrum of Sgr A East}

\begin{document}
\centerline{Submitted to the Editor of the Astrophysical Journal Letters.}
\vskip 0.5in
\title{A Self-Consistent Model for the Broad-band Spectrum\\
       of Sgr A East at the Galactic Center}

\author{Fulvio Melia\altaffilmark{1}$^{*\dag}$, Marco Fatuzzo$^*$,
Farhad Yusef-Zadeh$^\ddag$ and Sera Markoff\altaffilmark{2}$^{*}$}
\affil{$^*$Physics Department, The University of Arizona, Tucson, AZ 85721}
\affil{$^{\dag}$Steward Observatory, The University of Arizona, Tucson, AZ 85721}
\affil{$^{\ddag}$Department of Physics and Astronomy, Northwestern University,
Evanston, IL 60208}


\altaffiltext{1}{Presidential Young Investigator.}
\altaffiltext{2}{NSF Graduate Fellow.}


\begin{abstract}
Sgr A East is a very prominent elongated shell structure surrounding (though off-centered
from) the Galactic nucleus.  Its energetics ($\sim 4\times 10^{52}$ ergs), 
based on the power required to carve out the radio synchrotron remnant within 
the surrounding dense molecular cloud, appear to be extreme compared to 
the total energy ($\sim 10^{51}$ ergs) released in a typical supernova (SN) explosion.
Yet it shares several characteristics in common with SN remnants (SNRs),
the most significant of which is the $\sim 0.1-10$ GeV $\gamma$-ray 
spectrum measured by EGRET, if we associate the Galactic center
source 2EGJ1746-2852 with this nonthermal shell.  We here show that the
highest-energy component in Sgr A East's spectrum, like that of SNRs, can be 
fitted with the $\gamma$-rays produced in $\pi^0$ decays.
Further, we demonstrate in a self-consistent manner that the leptons released 
in the associated $\pi^\pm$ decays produce an $e^\pm$ distribution that
can mimic a power-law with index $\sim 3$, like that inferred from the VLA
data for this source.  These relativistic electrons and positrons also
radiate by bremsstrahlung, and inverse Compton scattering with the intense
IR and UV fields from the nucleus.  We show that the overall emissivity 
calculated in this way may account for Sgr A East's broadband spectrum
ranging from GHz frequencies all the way to TeV energies, where Whipple has thus
far set an upper limit to the flux corresponding to a $2.5\sigma$ signal 
above the noise. 

\end{abstract}




%

\section{Introduction}
Sgr A East is an elliptical shell structure elongated along the
Galactic plane with a major axis of length $10.5$ pc and a center
displaced from the apparent dynamical nucleus, Sgr A West, by $2.5$
pc in projection toward negative Galactic latitudes.  The latter 
is a ($\sim 1-2$ pc) three-arm spiral configuration of ionized gas 
(Ekers et al. 1983; Lo \& Claussen 1983) that engulfs a compact nonthermal
radio source, Sgr A*, having a dimension between 1.1 AU and 0.1 AU
(Morris \& Serabyn 1996).  The bulk of the thermally emitting Sgr A 
West appears to be located in front of Sgr A East (Yusef-Zadeh \& 
Morris 1987; Pedlar et al. 1989), whose exact distance behind the 
nucleus is not known, but a number of arguments suggest that it is 
positioned very close to the Galactic center (e.g., G\"usten \& 
Downes 1980;  Goss et al. 1989; Yusef-Zadeh et al. 1998).
Thus, if present, the massive black hole (presumably associated with 
Sgr A*) lies within the Sgr A East shell, but at one extreme end of 
its elongated structure.

Sgr A East may be a supernova remnant (SNR), but its inferred energetics
($\sim 4\times 10^{52}$ ergs;  Mezger et al. 1989) and size appear to be 
extreme and have generated some uncertainty regarding this interpretation.  
The explosion that produced Sgr A East may have been the tidal disruption 
of a main sequence star whose trajectory took it within ten 
Schwarzschild radii of the central object (Khokhlov \& Melia 1996).  
However, regardless of what actually caused the initial explosion, recent 
observations of this region at 1720 MHz (the transition
frequency of OH maser emission) have revealed the presence of several
maser spots on the remnant's boundary (Yusef-Zadeh et al. 1996).  This is important
because OH masers can be effective tracers of shocks produced at the 
interface with molecular clouds, particularly for supernova remnants
(Frail et al. 1994; Wardle, Yusef-Zadeh \& Geballe 1998).  
Radio continuum observations of
this object at $\lambda\lambda 20$ and $6$ cm (Pedlar et al. 1989)
reveal that these sites are also nonthermal emitters, but if this is due to synchrotron, 
then a relativistic electron distribution with $dN/dE\sim E^{-\alpha}$ is required, 
where $\alpha\sim 2.5-3.3$. This is rather peculiar for several 
reasons.  First, such a particle spectrum is unlike that 
($\alpha\sim 2.0-2.4$; Jones \& Ellison 1991) thought to be produced
directly in shock acceleration.  
Second, this situation appears to be different from that of other SNRs 
in which $\alpha\sim 2.2-2.4$ (e.g., Gaisser et al. 1998).

This latter distinction is very noticeable in view of
the kinship between Sgr A East and other SNRs suggested by EGRET observations
(see e.g., Green 1995; Esposito et al. 1996;
Saken et al. 1992; Lessard et al. 1997). The 
Galactic center is very complex and the $\gamma$-rays from this
region may be produced by other means.  For example, the EGRET
$\gamma$-ray source 2EGJ1746-2852 may be the black hole candidate
Sgr A* (Markoff, Melia \& Sarcevic 1997; Mahadevan et al. 1997) or 
it may be due to a process
different than what is producing the $\gamma$-ray emission in 
the SNRs (Melia, Yusef-Zadeh \& Fatuzzo 1998;  see below).
Our goal in this {\it Letter} is to demonstrate in a self-consistent manner
how the radio through TeV spectrum of Sgr A East may be produced with
a single process, and how it can be placed in context with the other
EGRET SNRs when one takes into account the differences between the ambient
medium surrounding Sgr A East and that of the other remnants. 

\section{Clues from the Observed Spectra of Sgr A East and the other SNRs}

EGRET has identified a central ($< 1^o$) $\sim 30$ MeV 
- $10$ GeV continuum source with luminosity $\approx 2\times 10^{37}$ ergs 
s$^{-1}$ (Mayer-Hasselwander et al. 1998).  Its
$\gamma$-ray flux does not vary down to the instrument sensitivity 
(roughly a $20\%$ amplitude), suggesting that it may be diffuse
(see, however, Markoff, Melia \& Sarcevic 1997).  In a recent study (Melia,
Yusef-Zadeh \& Fatuzzo 1998), we showed that the $\gamma$-ray characteristics
of 2EGJ1746-2852 may be understood as arising from inverse Compton scatterings 
between the power-law e$^-$ (inferred from the radio data) and the UV and IR
photons from the nucleus.  This is facilitated by the unique environment
at the Galactic center, where the inner $1-2$ pc region is a source of
intense (stellar) UV and (dust-reprocessed) far-IR radiation that bathes
the extended synchrotron-emitting structures.

However, this analysis left several questions unanswered.  For example, 
it is rather suspicious that the Galactic center
$\gamma$-ray source has a similarly distinctive bump at $\sim 0.1-1$ GeV as
the five SNRs detected by EGRET (see, e.g., Esposito et al. 1996).  
In their detailed calculations of the emission in the SNR shells of IC 443 and 
$\gamma$ Cygni, Gaisser,
Protheroe and Stanev (1998) included the contribution to IC from microwave background 
radiation, the Galactic IR/optical background radiation and the IR 
resulting from shock-heated dust. At least in these SNRs, the dominant
contributions to the $\gamma$-ray spectrum come from bremsstrahlung
(by relativistic electrons colliding with the ambient proton distribution)
and pion decay, the latter constituting the aforementioned $\sim 70$ MeV
characteristic bump.  Thus, if 2EGJ1746-2852 is the $\gamma$-ray counterpart
to the radio emitting shell in Sgr A East, one might expect that
its $\gamma$-ray bump could also be due to pion decays.
At the same time, an observationally significant
pion production rate in Sgr A East also offers the exciting possibility that 
the $e^+e^-$ pion decay byproducts may produce the $\alpha\sim 3$ power-law
distribution inferred from Sgr A East's radio spectrum, and these particles
then also self-consistently account for the underlying $\gamma$-ray spectrum 
via either bremsstrahlung, or the IC scattering of the central IR and UV photons. 

We work with the hypothesis that the EGRET $\gamma$-ray bump
is due primarily to $\pi^0$ decays ($\pi^0\rightarrow\gamma\gamma$), 
these being produced in $pp$ scatterings between ambient protons and 
a relativistic population accelerated by shocks within the Sgr A East shell.
In this process, a multiplicity of pions is produced, the charged
members of which decay leptonically ($\pi^\pm\rightarrow \mu^\pm\nu_\mu$, 
with $\mu^\pm\rightarrow e^\pm\nu_e \nu_\mu $).  We first find the physical
conditions required to produce this $\gamma$-ray bump, and then use
the associated electrons and positrons produced in the pion decays to
calculate the radio spectrum.  Fitting the synchrotron emission to the
VLA data (see Fig. 3) is then a matter of adjusting just one unknown,
viz. the average magnetic field $B$ in the gyration region, though this
is expected to be close to equipartition.  

A full description of our $pp$ scattering algorithm is given elsewhere
(Markoff, Melia \& Sarcevic 1997, 1998).  Briefly, we assume that the
accelerated protons are accelerated at the shock with a rate of
$\dot{\rho}_p(E_p)=\rho_oE_p^{-\alpha_p}$ cm$^{-3}$ s$^{-1}$ GeV$^{-1}$.  The
steady-state proton spectral index is thought to lie in the range
of $2.0-2.4$ (Jones \& Ellison 1991), and is determined from a diffusion loss
treatment.  Because the protons are ultrarelativistic, the leading order nucleons 
produced will continue contributing to the spectrum via multiple collisions 
in an ensuing cascade, until they lose enough of their energy to rejoin the 
ambient plasma. The procedure for calculating the pion production rate is 
complicated by the energy dependence of the 
pion multiplicity and the $pp$ scattering cross section.  
Fortunately, uncertainties in the $pp$ scattering
are likely to be smaller than $\sim 20-30\%$, within the possible
errors of the astrophysical components, such as the source size.
We calculate the steady-state proton distribution resulting from the balance of
injection from the shock, cascade collisions producing secondary
particles which feed back into the system, and synchrotron and inverse
Compton losses.  We note that proton-photoproduction
processes are here relatively insignificant, since the maximum energy
attained by the relativistic protons in the face of energy losses due
to $pp$ scatterings and other processes is about $5\times 10^6$ GeV,
close to the pair creation threshold for protons scattering off UV
radiation (see below).

In Figure 1, we show the $\gamma$-ray spectrum produced from 
$\pi^0$ decays for two different cases, characterized by their
respective values of $\alpha_p$ and the ambient proton density $n_p$.
Of course, the full spectrum will be the sum
of this component and those arising from bremsstrahlung and inverse
Compton scattering, which we consider below.  However, we can already
see that the fit to the $\gamma$-ray data between $\sim 100$ MeV
and $10$ GeV is rather robust over a range
of conditions.  What does change from case to case is the energy
content in the relativistic proton population, which is related to the
efficiency introduced above.  So, for example, when the ambient proton
density $n_p$ is low (Case 2), the removal rate of protons from the 
relativistic distribution is smaller than that in Case 1, resulting in a
higher equilibrium density of relativistic protons. This is why the
energy content of these particles is much higher in Case 2 than in Case 1,
accounting for roughly $77\%$ and $1\%$ of a typical SN burst of
$10^{51}$ ergs for these two situations, respectively. 

The electron distribution resulting from these two Cases is shown
in Figure 2.  The $e^-$ and $e^+$ distributions can mimic a power-law with a spectral
index $\sim 3$ very well, consistent with the value implied by the
radio spectrum of Sgr A East.  
Apparently, the leptons produced in $\pi^\pm$ decays in this source
are numerically dominant over those accelerated directly within shocks.
In fact, assuming the same spectral index for the electrons and protons,
the shock would have to accelerate roughly $1,200$ as many
of the former compared to the latter at any given energy
in order to account for the synchrotron emissivity
without the decay leptons.  This imbalance comes about because (1)
each relativistic proton produces an average of $20-30$ electrons
and (2) a shock accelerated electron distribution would contain a
much larger contribution from low energy electrons than the distribution
shown in Figure 2, which shows a deficiency at small $\gamma$.

These relativistic electrons are immersed in an intense source of
UV and IR radiation from the central $1-2$ parsecs.  
The Circumnuclear Disk (CND) is a powerful source ($\approx 10^7\:L_\odot$) of
mid to far-infrared continuum emission with a dust temperature of
$\approx 100$ K (e.g., Telesco et al. 1996; Davidson et al. 1992).
This radiation is due to reprocessing by warm dust that has absorbed 
the same power in the UV (Becklin, Gatley and Werner 1982;
Davidson et al. 1992). Models of the photodissociation regions in the CND require
an incident flux ($6$ eV $< h\nu < 13.6$ eV) of $10^2$--$10^3$
erg cm$^{-2}$ s$^{-1}$ (Wolfire, Tielens \& Hollenbach, 1990), implying
a total UV luminosity of about $2\times 10^7\;L_\odot$. 
This radiation field has a specific
photon number density per solid angle $n_{ph}^{tot}(\varepsilon)\equiv
n_{ph}^{UV}(\varepsilon)+n_{ph}^{IR}(\varepsilon)$, where
$n_{ph}^{UV}(\varepsilon) = N_0^{UV}(2\varepsilon^2/h^3 c^3)
(\exp\{\varepsilon/kT^{UV}\}-1)^{-1}$, and $n_{ph}^{IR}(\varepsilon) =
N_0^{IR}(2\varepsilon^3/h^3 c^3)(\exp\{\varepsilon/kT^{IR}\}-1)^{-1}$.
Here, $\varepsilon$ is the lab-frame photon energy and $T^{UV}$ and
$T^{IR}$ are, respectively, the temperature (assumed to be $30,000$ K)
of the stellar UV component and of the reprocessed CND radiation, which
is assumed to peak at $50\mu$m, corresponding to a characteristic temperature
$T^{IR}\approx 100$ K.  The normalization
constants $N_0^{UV}$ and $N_0^{IR}$ incorporate the dilution in photon
number density as the radiation propagates outwards from the central
core.  This is calculated assuming that the radiation
emanates from a sphere of radius $\approx 1$ pc for the UV, and that
the IR luminosity is $10^7\,L_\odot$ from a disk with radius $\approx 2$ pc.

The calculation of the inverse Compton (X-ray and $\gamma$-ray) emissivity
for this known $e^+e^-$ population bathed by the IR and UV radiation
is carried out according to what is now a standard procedure (see, for example,
Melia, Yusef-Zadeh \& Fatuzzo 1998). The upscattered radiation is emitted 
isotropically from within a volume 
$V\sim 250$ pc$^3$ in Sgr A East, corresponding to a shell
with radius $R\approx 5$ pc and thickness $\Delta R\approx 1$ pc.  The centroid
of this structure is assumed to lie at $7$ pc behind the nucleus.  
The remaining continuum component of importance is bremsstrahlung radiation
resulting from the interaction between the relativistic leptons and the ambient 
(fixed target) nuclei (see Koch \& Motz 1959).  
Thus, the overall spectrum from Sgr A East is a superposition of the $\gamma$-rays
from $\pi^0$ decays, synchrotron radiation by the relativistic leptons produced
during the decay of the charged pions, bremsstrahlung emission by these electrons
and positrons, and their Comptonization of the IR and UV radiation 
from the central $1-2$ pc.

\section{Discussion}

The key result of our calculations is displayed in Figure 3 for Case 2 above.
The inferred magnetic field is $\approx 10^{-5}$ Gauss, which is within
a factor of $2$ of the equipartition value of $B$ in Sgr A East, assuming a radio
luminosity of $1.8\times 10^{35}$ ergs s$^{-1}$ arising from relativistic leptons 
gyrating in a magnetic field over a volume of $250$ pc$^{3}$.
The ASCA measurement is an upper limit, since it does not distinguish between
thermal and nonthermal emission from this region.  
Our predicted hard X-ray spectrum is also consistent with Ginga observations, which 
show evidence of a hard tail in the diffuse X-ray spectrum from the inner region 
of the Galaxy (Yamasaki et al. 1996), and with the current OSSE upper limits
(Purcell et al. 1997). 
At the highest energy measurement to date, a 270 minute exposure by the Whipple 
observatory produced a small excess ($2.5\sigma$) of $> 2.0$ TeV emission from 
within the EGRET error box of 2EG J1746-2852, yielding an integrated flux upper 
limit of $0.45\times 10^{-11}$ cm$^{-2}$ s$^{-1}$ (Buckley et al. 1997).  
Although this signal is at best only $2.5\sigma$ above the noise, our calculated
spectrum would be consistent with a detection at this level.  

Case 1 would not produce an acceptable spectrum because the larger ambient proton
density ($n_p=10^3$ cm$^{-3}$) results in a significantly higher lepton
bremsstrahlung flux.  The
spectral shape would then also not be a good fit to the data since the $\gamma$-rays
from $\pi^0$ decays would be swamped by the other radiative components.  In
fact, we found it very difficult to fit this broadband spectrum with any ambient
density in excess of $\sim 10$ cm$^{-3}$.  This raises an interesting issue,
since the energy content in the relativistic protons (see above) is then a large 
fraction ($\sim 0.5-0.8$) of the energy in a ``standard'' SN burst ($\approx 10^{51}$ ergs).
On the other hand, this relativistic energy content would constitute only
$\sim 1-2\% $ of the available energy if the burst that produced the Sgr A East
remnant was $\sim 4\times 10^{52}$ ergs, as suggested by other observations
(see above).  Our results may therefore be somewhat consistent with the supposition
that Sgr A East was produced not by a single, standard supernova explosion, but 
rather by several such events, or a single more catastrophic incident, such as the 
tidal disruption of a star. 
In both Cases 1 and 2 we adopted a value of $\alpha_p$ within a narrow range
close to $2$.  We found our fit to be very sensitive to $\alpha_p$.  Even
with the large error bars associated with the VLA points, a proton distribution
steeper than this produces a steeper lepton distribution, whose
synchrotron emissivity does not fit the radio data. A shallower
proton index is, of course, uncomfortably below the theoretical range.

If the $\gamma$-rays detected by EGRET are indeed produced in this
fashion, it seems inevitable to us that the upscattered IR and UV photons 
should result in a significant intensity at intermediate (i.e.,
$\sim 10-100$ keV) energies, as indicated in Figure 3.  This flux 
density ($\sim 10^{-3}$ photons cm$^{-2}$ s$^{-1}$ MeV$^{-1}$) 
may be above the sensitivity limit (at $\sim 10^{-7}$ photons cm$^{-2}$ s$^{-1}$ 
MeV$^{-1}$ for a point source) of UNEX, a proposed balloon flight instrument 
(Rothschild 1998).  

Other than the overall energetics, it appears that the main distinction
between Sgr A East and the SNRs is the intensity of the
external radiation upscattered by the relativistic electrons and positrons.  Whereas
the ambient soft photon intensity in the latter is too low to
produce an observationally significant spectral component compared to
bremsstrahlung (Gaisser, Protheroe \& Stanev 1998), the IR and UV radiation
from the Galactic center does produce a dominant broadband continuum
when scattering with the relativistic leptons in Sgr A East.

{\bf Acknowledgments} This research was partially supported by NASA under grant 
NAGW-2518 and by an NSF Graduate Fellowship.

{}

%
%

\newpage


\begin{figure}[thb]\label{fig1.ps}
{\begin{turn}{-90}
\epsscale{0.85}
\plotone{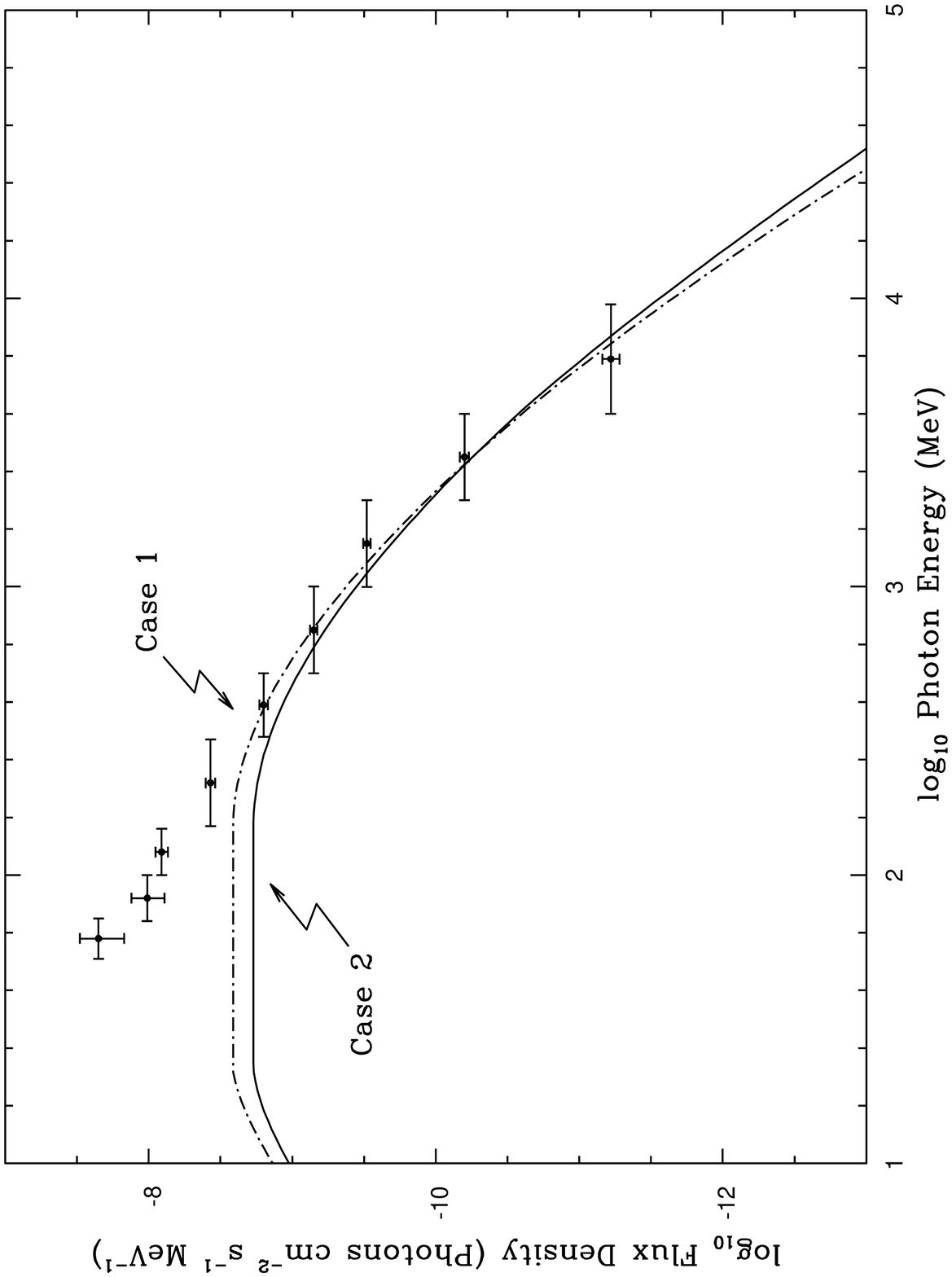}
\end{turn}}
\caption{The $\gamma$-ray spectrum due to pion decays in
the Sgr A East shell.  The two cases shown here represent the range of physical
conditions within which a good fit may be obtained simultaneously for the
EGRET and VLA data (see text).  Case 1: $n_p=10^3$ cm$^{-3}$, $\alpha_p=2.2$.
Case 2: $n_p=10$ cm$^{-3}$, $\alpha_p=2.0$.  In both cases, the shell is assumed
to have a radius of $5$ pc and a width of $1$ pc.  In steady state, the
relativistic (shock accelerated) protons constitute a total energy of
$8.8\times 10^{48}$ ergs for Case 1, and $7.7\times 10^{50}$ ergs for Case 2.
The EGRET data points are taken from Mayer-Hasselwander et al. (1998).}
\end{figure}

\clearpage
\begin{figure}[thb]\label{fig2.ps}
{\begin{turn}{0}
\epsscale{1.00}
\plotone{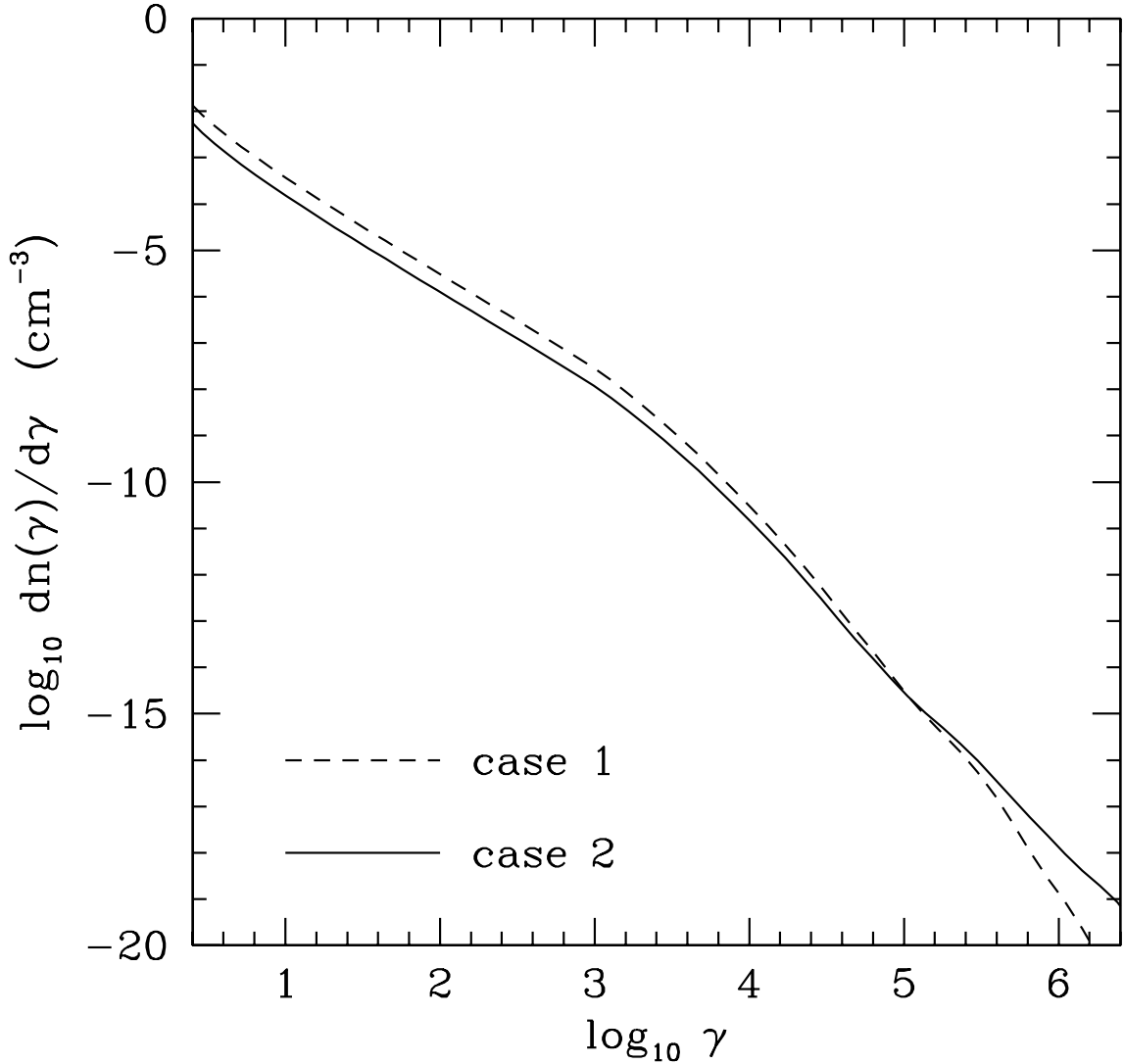}
\end{turn}}
\caption{The steady state electron distribution resulting from pion decays
for the same population that produces the spectra shown in Fig. 1.  Cases 1 and 2
correspond to those defined in that figure.  The positron population is the same
as that shown here for the electrons.  The steady state results from a combination
of pion decays and an energy loss due to synchrotron, and inverse Compton scatterings
with the IR and UV photons from the nucleus.  The significance of this plot is
that the lepton distribution produced in this way can mimic a power-law with
a spectral index $\sim 3$ rather than the canonical value of $\sim 2$ produced in
standard shock acceleration.  Only an electron-positron distribution of this type
can account for the observed synchrotron spectrum at GHz frequencies.}
\end{figure}

\clearpage
\begin{figure}[thb]\label{fig3.ps}
{\begin{turn}{-90}
\epsscale{0.85}
\plotone{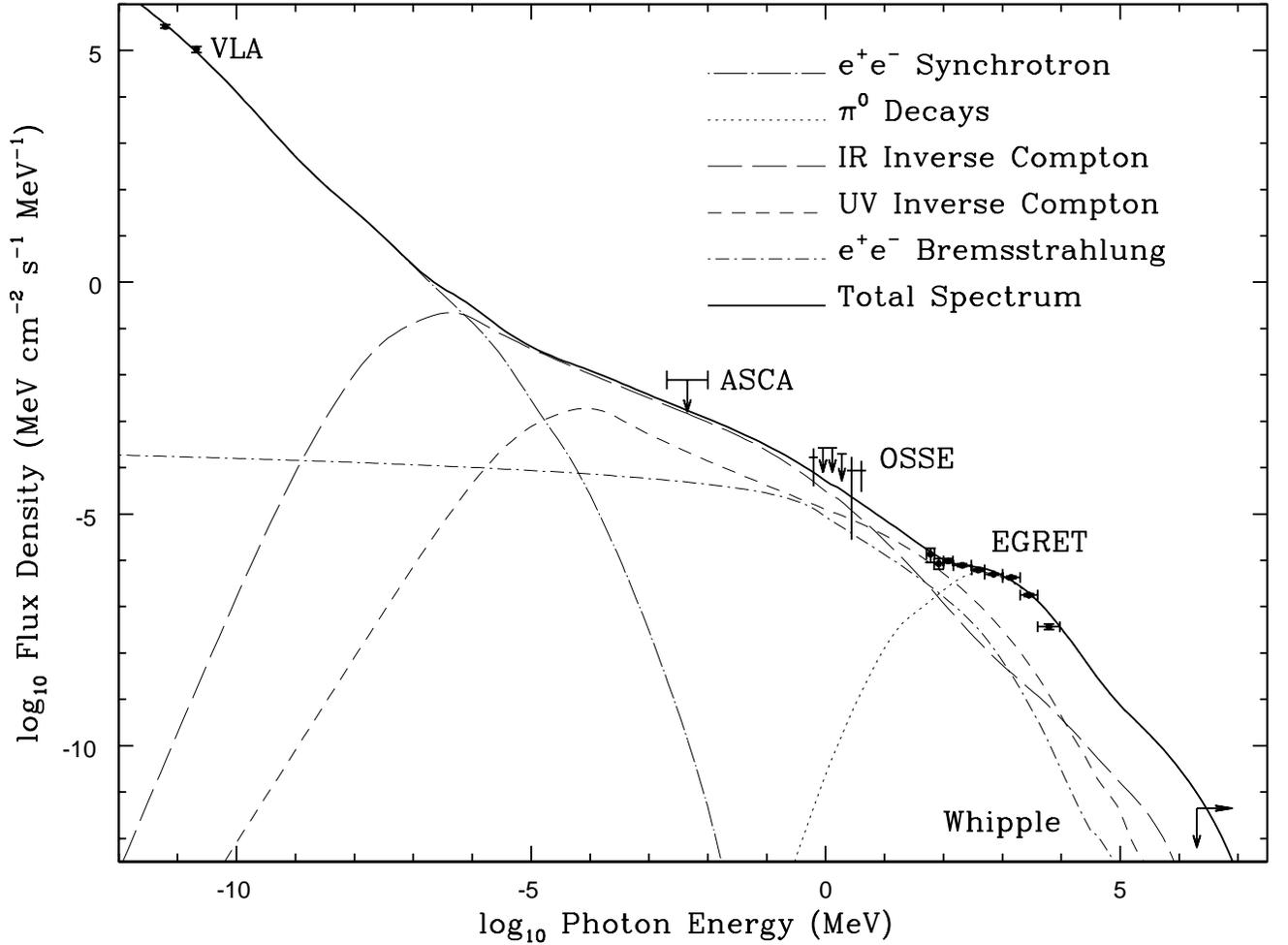}
\end{turn}}
\caption{The broadband spectrum calculated self-consistently using the particle
decay products from the fit of Case 2 in Fig. 1, i.e., each of the spectral components
is calculated using the electron (and corresponding positron) distribution shown for
Case 2 in Fig. 2.  Once the EGRET data are fit with the pion-decay $\gamma$-rays,
the rest of the spectrum shown here is fixed.  The magnetic field inferred from the
VLA data is $B(\hbox{\rm Sgr A East})\approx 1.1 \times 10^{-5}$ G.  The data included in this
plot are from Pedlar et al. (1989) (VLA), Koyama et al. (1996) (ASCA), Purcell
et al. (1997) (OSSE), Mayer-Hasselwander et al. (1998) (EGRET), and Buckley et al. (1997) (Whipple).}
\end{figure}

\end{document}